\documentclass[11pt, letterpaper]{amsart}
\usepackage{textcomp}

\theoremstyle{definition}

\theoremstyle{remark}

\numberwithin{equation}{section}
\newcommand{\dbra}[1]{[\kern-0.15em[ #1 ]\kern-0.15em]}
\newcommand{\dbraco}[1]{[\kern-0.15em[ #1 [\kern-0.15em[}
\newcommand{\dbraoc}[1]{]\kern-0.15em] #1 ]\kern-0.15em]}

\begin{document}

\title[Arbitrage strategy]{Arbitrage strategy}%

\author{Constantinos Kardaras}%
\address{Constantinos Kardaras, Mathematics and Statistics Department, Boston University, 111 Cummington Street, Boston, MA 02215, USA.}%
\email{kardaras@bu.edu}%

%\thanks{}%
%\subjclass{60H05 · 60H30 · 91B28}%
%\keywords{}%

%\date{\today}%
%\dedicatory{}%
%\commby{}%
% ----------------------------------------------------------------
\begin{abstract}
An \textsl{arbitrage strategy} allows a financial agent to make certain profit out of nothing, i.e., out of zero initial investment. This has to be disallowed on economic basis if the market is in equilibrium state, as opportunities for riskless profit would result in an instantaneous movement of prices of certain financial instruments. The principle of not allowing for arbitrage opportunities in financial markets has far-reaching consequences, most notably the option-pricing and hedging formulas in complete markets.
\end{abstract}

\maketitle

% ----------------------------------------------------------------

It is difficult to imagine of a normative condition that is more widely-accepted and unquestionable in the minds of anyone involved in the field of Quantitative Finance other than the \emph{absence of arbitrage opportunities} in a financial market. Put plainly, an \textsl{arbitrage strategy} allows a financial agent to make certain profit out of nothing, i.e., out of zero initial investment. This has to be disallowed on economic basis if the market is in equilibrium state, as opportunities for riskless profit would result in an instantaneous movement of prices of certain financial instruments.

\smallskip

Let us give an illustrative example of an arbitrage strategy in the foreign exchange market, commonly called the \emph{triangular arbitrage}. Suppose that Kate, in London, is buying\footnote{All the prices we are referring to in this example are \textsl{bid} prices of the currencies involved.} the U.S. dollar for $\textrm{\texteuro} 0.685$. Tom, in San Francisco, is buying Japanese yen for $\$ 0.009419$. Finally, Toru, in Tokyo, is buying one euro for $\yen 155.02$. All these transactions are supposed to be able to occur at the \emph{same time}. There is something particular in the situation just-described --- something that could allow you to make riskless profit. Let us see how. You borrow $\$ 10,000$ from your rich aunt Clara and tell her you will return the money in a matter of minutes. First, you approach Mary and change all your dollars to euros. This means that you will get $\textrm{\texteuro} 6,850$. With the euros in-hand, you contact Toru and change them into yen --- you will get $\yen (6,850
\times 155.02) = \yen 1,061,887$. Finally, you call Tom, wire him all your yen and change them back to dollars, which gets you $\$ (1,061,887 \times 0.009419) \equiv \$ 10,001. 91$. You give the $\$10,000$ back to you aunt Clara as promised, and you have managed to create $\$1. 91$ out of thin air.

Although the above example is over-simplistic, it gives a clear idea of what arbitrage is: a position on a combination of assets that requires zero initial capital and results in a profit with \emph{no} risk involved. Let us now walk a step further and see what will happen under the situation of the preceding example. As more and more investors become aware of the discrepancy between prices, they will all try to use the same smart strategy that you used for their benefit. Everyone will be trying to exchange U.S. dollars for euros in the first step of the arbitrage, which will drive Mary to start buying U.S. dollar for less than $\textrm{\texteuro} 0.685$, as a result of high demand for the euros she is selling. Similarly, Tom will start buying Japanese yen for less than $\$ 0.009419$ and Toru will be buying euro for less than $\yen 155.02$. Very soon, the situation will be such that nobody is able to make a riskless profit anymore.

\smallskip

The economic rationale behind asking for non-existence of arbitrage opportunities is exactly based on the discussion of the previous paragraph. If arbitrage opportunities were present in the market, a multitude of investors would try to take advantage of them simultaneously. Therefore, there would be an almost instantaneous move of the prices of certain financial instruments as a response to a supply-demand imbalance. This price movement will continue until there is no longer any opportunity for riskless profit available.

It is important to note that the above, somewhat theoretical, discussion does \emph{not} imply that arbitrage opportunities \emph{never} exist in practice. On the contrary, it has been observed that opportunities for some, albeit usually minuscule, riskless profit appear frequently as a consequence of the huge amount of distant geographic trading locations, as well as a result of the numerous financial products that have sprung and are sometimes interrelated in complicated ways. Realizing that such opportunities exist is a matter of rapid access to information that a certain group of investors, so-called \textsl{arbitrageurs}\footnote{See \cite{citeulike:2550005} for a classification of investor types.}, has. It is rather the \emph{existence} of  arbitrageurs acting in financial markets that ensures that when arbitrage opportunities exist, they will be fleeting.

\smallskip

The principle of not allowing for arbitrage opportunities in financial markets has far-reaching consequences and has immensely boosted research in Mathematical Finance. The ground-breaking papers of F. Black and M. Scholes \cite{citeulike:202505} and R. Merton \cite{Merton73}, published at 1973\footnote{For this work, the authors were awarded the Nobel prize in Economics in 1997.}, were the first instances explaining how absence of arbitrage opportunities leads to rational pricing and hedging formulas for European-style options in a geometric Brownian motion financial model.
%\footnote{For historical perspectives regarding option pricing and hedging, see [eqf01.004], [eqf01.009], [eqf01.010], [eqf01.014]. For a more thorough quantitative treatment, check [eqf04.003].}
This idea was consequently taken up and generalized by many authors, and has lead to a profound understanding of the interplay between the economics of financial markets and the mathematics of stochastic processes, reaching deep results.
%--- see [eqf04.002], [eqf04.003], [eqf04.007], [eqf04.013] for some amazing developments on this path.

\smallskip

We close the discussion of arbitrages on a funny and light note. Such is the firm belief on the principle of not allowing for arbitrage opportunities in financial modeling, that even jokes have been created in order to substantiate it further. We quote directly from Chapter 1 of \cite{MR2200584}, which can be used as an excellent, though more mathematically advanced, introduction to arbitrage.
\begin{quote}
\emph{A professor working in Mathematical Finance and a normal\footnote{Is this bold distancing from normality of Mathematical Finance professors, clearly implied from the authors of \cite{MR2200584}, a decisive step towards illuminating the perception they have of their own personalities? Or is it just a gimmick used to add another humorous ingredient to the joke? The answer is left for the reader to determine.} person go on a walk and the normal person sees a \texteuro 100 bill lying on the street. When the normal person wants to pick it up, the professor says: ``Don't try to do that. It is absolutely impossible that there is a \texteuro 100 bill lying on the street. Indeed, if it were lying on the street, somebody else would have picked it up before you''.}
\end{quote}

% ----------------------------------------------------------------
\bibliographystyle{siam}
\bibliography{arbitrage_strategy}

\begin{thebibliography}{1}

\bibitem{citeulike:202505}
{\sc F.~Black and M.~Scholes}, {\em The pricing of options and corporate
  liabilities}, The Journal of Political Economy, 81 (1973), pp.~637--654.

\bibitem{MR2200584}
{\sc F.~Delbaen and W.~Schachermayer}, {\em The mathematics of arbitrage},
  Springer Finance, Springer-Verlag, Berlin, 2006.

\bibitem{citeulike:2550005}
{\sc J.~C. Hull}, {\em Options, Futures, and Other Derivatives (7th Edition)},
  {Prentice Hall}, May 2008.

\bibitem{Merton73}
{\sc R.~C. Merton}, {\em {Theory of Rational Option Pricing}}, Bell Journal of
  Economics, 4 (1973), pp.~141--183.

\end{thebibliography}
\end{document}